\documentclass[aps,preprint,groupedaddress,showpacs]{revtex4}
\usepackage{bm,amsfonts,amssymb,amsmath}
\usepackage{graphicx}

\begin{document}

\title{Adsorption of a binary mixture of monomers with
nearest-neighbour cooperative effects}

\author{A. Prados}
\email{prados@us.es}
\author{J. J. Brey}
\email{brey@us.es}

\affiliation{F\'{\i}sica Te\'orica, Universidad de Sevilla,
Apartado de Correos 1065, E-41080 Sevilla, Spain}

\date{\today}

\begin{abstract}
A model for the adsorption of a binary mixture on a
one-dimensional infinite lattice with nearest neighbour
cooperative effects is considered. The particles of the two
species are both monomers but differ in the repulsive interaction
experienced by them when trying to adsorb. An exact expression for
the coverage of the lattice is derived. In the jamming limit, it
is a monotonic function of the ratio between the attempt
frequencies of the two species, varying between the values
corresponding to each of the two single species. This is in
contrast with the results obtained in other models for the
adsorption of particles of different sizes. The structure of the
jamming state is also investigated.
\end{abstract}


\maketitle

\section{\label{intro}Introduction}

The adsorption of particles on a solid substrate is a process
occurring in many natural phenomena and that, consequently, has
been extensively studied both experimentally and theoretically
\cite{Ev93}. The simplest kind of models for adsorption are those
defined in terms of random sequential adsorption (RSA) processes,
in which particles of a given kind are adsorbed at random
positions on the substrate. When the adsorption rates depend on
the state of the neighborhood of the position at which they take
place, the process is known as cooperative sequential adsorption
(CSA). A lattice version of the CSA process is the monomer filling
with nearest-neighbor (NN) cooperative effects \cite{Ke63,Ev93}.
These lattice models have been extensively used to study the
monolayer growth of monodisperse particles. In particular, exact
analytical solutions have been derived for the one-dimensional
case \cite{Ev93}.

The deposition of a mixture of different kind of particles has
also received some attention
\cite{TyS89,BayPr91,MyJ92,BAKyK96,Pa99,Bo01,Bo92,KSByK02a,KSByK02b},
but the progress made in its understanding is small as compared
with the one-component problem. Here, a one-dimensional lattice
model for a binary mixture of monomers with NN cooperative effects
is considered and its analytical solution is found. Besides their
theoretical interest, one-dimensional models are also useful for
the description of some polymer reactions \cite{Ev93}. A
controversial point in the adsorption of a mixture is whether it
covers the substrate more or less efficiently than either of the
species separately. Some results seemed to indicated than the
former was true for lattice models \cite{BayPr91,Bo92}, while the
latter applied for continuum models \cite{Bo01}. Nevertheless,  a
(continuum) random car parking model has been found to be
consistent with the lattice models in this respect
\cite{KSByK02a,KSByK02b}. For the lattice model considered here,
it will be shown than the jamming coverage of the mixture is
always between the coverage associated with each of the two
components. This appears to be a direct consequence of the NN
cooperative effects taken into account in the model.

It is interesting to note that lattice models with NN cooperative
effects have also been considered in the context of Ising models.
They were introduced by Fredrickson and Andersen
\cite{FyA84,FyA85,FyB86} to study structural relaxation in glassy
systems. In these models, a given spin can flip only if its
nearest neighbors are in a certain subset of all their possible
configurations and, because of this, they were termed
``facilitated'' Ising models. In particular, in the so-called
nSFM, one spin can flip only if at least $n$ of its NN spins are
orientated against the external field. It is easily verified that
the zero temperature limit of these models is the spin
representation of the hole-particle (monomer) models mentioned
above, i.e. both are related by a single change of variables.
Interestingly, the one-dimensional 1SFM has also been used as a
model to describe compaction proceses in vibrated granulates
\cite{BPyS99,PByS00,PyB03}.

In the model considered in this paper, the two kind of particles
composing the mixture differ in their dynamics. Particles of one
of the species obey RSA with  NN exclusion or blocking
\cite{Ev93}, i.e. they can occupy an empty site with the
constraint that both of its NN must be empty. On the other hand,
adsorption of one particle of the other species on an empty site
only requires that one of the NN be empty. Then, the dynamics of
both kind of particles is cooperative, but the ``facilitation''
rules are different. A possible interpretation of this form of
competitive adsorption is that particles are of the same size, but
the effective repulsive interactions coming from the adsorbed
particles are much stronger for one species than for the other.
Then, the repulsive effect depends on the kind of particle trying
to adsorb and on whether the NN sites are occupied or empty, but
not on the type of particle actually occupying them.

In the next Section, the model will be formulated in detail and
the master equation describing its dynamics will be written down.
The dynamics of the adsorption processes is studied in Sec.
\ref{dynamics}, where explicit analytical expressions for the time
evolution of the densities of each component are derived.  For
very large times, the system gets stuck in a jammed configuration,
where no more adsorption events are possible. The asymptotic
values of the densities of both species, as well as the total
coverage, are also calculated as a function of the adsorption rate
constants of the two species. It is found that the maximum
asymptotic total coverage is always smaller than the one
corresponding to the least repulsive particles. In order to get a
deeper understanding of the stationary, jammed, state reached by
the lattice in the long time limit, some correlation functions are
analysed in Sec.\ \ref{corr}. The jammed states are characterised
by having all the holes isolated, i.e., surrounded by two
particles. Therefore, we will study the structure of the lattice
around these isolated holes, analysing the probability of finding
them between all the possible configurations of their NN sites.
Section \ref{conc} contains a short summary of the main results
and conclusions of the paper. Finally, some calculations are
presented in an Appendix.

\section{\label{model}The model}

We consider two kinds of particles, A and B, that can be randomly
adsorbed on a one-dimensional lattice. The adsorption processes on
an empty site are restricted by the configuration of its NN, as
follows:

(i) Particles A can be adsorbed on an empty site only if at least
one of its NN is also empty.

(i) Adsorption of a particle B on a site requires that both of the
NN are empty.

Let us introduce two occupation numbers $n_{i}$ and $m_{i}$ for
each site $i$ of the lattice. If there is a particle A at site
$i$, it is $n_{i}=0$, otherwise $n_{i}=1$. Similarly, $m_{i}=1$ if
there is a particle B at site $i$, and $m_{i}=0$ if there is not.
Of course, since there can be at the most one particle at each
site, it is $(1-n_{i})(1-m_{i})=0$ for all $i$. A lattice
configuration is specified by the sets of occupation numbers of
the two species $\{ {\bm n},{\bm m} \}$, where ${\bm n} \equiv \{
n_{i} \}$ and ${\bm m} \equiv \{  m_{i} \}$.

The dynamics of the system is defined as a Markov process, so it
is enough to specify the master equation for the probability
distribution of the configurations, $p({\bf n},{\bf m},t)$. As
said above, the elementary processes are the adsorption of
particles A and B on an empty site $i$. In these events, the
configuration of the system changes from $ ({\bf n},{\bf m})$ to
$(R_{i}{\bf n},{\bf m})$ or to $({\bf n},R_{i}{\bf m}$), depending
on whether the adsorbed particle corresponds to species A or B.
Here $R_{i}$ is the operator changing the occupation number
$n_{i}$ or $m_{i}$ into $1-n_{i}$ or $1-m_{i}$, respectively.
Thus, the master equation reads
\begin{eqnarray}
\frac{d}{dt} p(\bm{n},\bm{m},t) & = & \sum_i \left[
W_i(\bm{n},\bm{m}|R_i\bm{n},\bm{m})
p(R_i\bm{n},\bm{m},t)-W_i(R_i\bm{n},\bm{m}|\bm{n},\bm{m})
p(\bm{n},\bm{m},t) \right] \nonumber \\
& + & \sum_i \left[ W_i(\bm{n},\bm{m}|\bm{n},R_i\bm{m})
p(\bm{n},R_i\bm{m},t)-W_i(\bm{n},R_i\bm{m}|\bm{n},\bm{m})
p(\bm{n},\bm{m},t) \right] \, . \label{2.1}
\end{eqnarray}
where
\begin{equation}
\label{2.2} W_i(R_i\bm{n},\bm{m}|\bm{n},\bm{m})=\frac{\alpha}{2}
n_i m_i \left( n_{i-1}m_{i-1}+n_{i+1}m_{i+1} \right) \, ,
\end{equation}
\begin{equation}
\label{2.3} W_i(\bm{n},R_i\bm{m}|\bm{n},\bm{m})=\beta n_{i-1}
m_{i-1} n_i m_i n_{i+1} m_{i+1} \, .
\end{equation}
The above transitions rates describe the adsorption of a particle
A and B on site $i$, respectively, consistently with the rules
$(i)$ and $(ii)$. The constant parameters $\alpha$ and $\beta$
characterise the attempt frequency for each kind of adsorption
process.

If $\alpha=0$ or $\beta=0$, the lattice will be occupied by
particles of only one kind, and the dynamics of the model reduces
to the one-component monomer filling with NN cooperative effects,
whose solution is known and discussed in  detail by Evans in Ref.\
\cite{Ev93}. When both rates are nonzero, we have the competitive
adsorption of the two species:  particles B need both of the NN
sites of a given hole to be also empty in order to adsorb, while
one empty NN site is enough for a particle A. In the corresponding
facilitated Ising picture \cite{FyA84,FyA85,FyB86}, if $\beta=0$
we have the one-dimensional 1SFM, while for $\alpha=0$ the
one-dimensional 2SFM is recovered.

We will restrict ourselves to translationally invariant states.
This requires the consideration of consistent boundary conditions,
e.g. periodic ones. In this case, the density of particles A at a
given time $t$ is given by
\begin{equation}
\label{3.1} \rho_{A}(t)= 1- \langle n_{i}\rangle_{t},
\end{equation}
where the angular brackets denote average with the probability
distribution $p({\bf n},m,t)$ and the right hand site does not the
depend on the site $i$ chosen. Similarly, the density of B
particles is
\begin{equation}
\label{3.1b} \rho_{B}(t)= 1- \langle m_{i} \rangle_{t}.
\end{equation}
The total coverage of the line $\theta (t)$ is then
\begin{equation}
\label{3.1c} \theta(t)=\rho_A(t)+\rho_B(t) \, .
\end{equation}
It verifies $ 0 \leq \theta (t) <1$, since it is impossible to
fully fill the line, as it will be discussed in detail latter on.
A particularly interesting property is the asymptotic coverage
$\theta_{J}$ at jamming, i.e., once the system gets stuck in a
configuration in which no more adsorption events are possible. In
the model considered here, this happens when all the holes are
isolated.  Formally, we can write
\begin{equation}
\label{3.1d} \theta_J=\lim_{t\rightarrow\infty} \theta (t) \, .
\end{equation}
It is useful to consider the set of moments
\begin{equation}
\label{3.2} F_r(t)=\langle n_i m_i n_{i+1} m_{i+1} \ldots n_{i+r}
m_{i+r} \rangle_t \, ,
\end{equation}
for all $r  \geq 0$. They give the probability of finding $r+1$
consecutive holes or, equivalently, the density of clusters of at
least $r+1$ empty sites. In particular, $F_{0}(t)$ gives the
density of holes and, therefore, it must be
\begin{equation}
\label{3.3} F_0(t)+\rho_A(t)+\rho_B(t)=1 \, .
\end{equation}
The above equation can also be written as
\begin{equation}
\label{3.4} \langle (1-n_i)(1-m_i) \rangle_t=0 \, ,
\end{equation}
expressing that a site can not be simultaneously occupied by a
particle A and a particle B. Besides, between each pair of
particles B there must be at least one empty site, due to the
adsorption rule for particles A. Therefore,
\begin{equation}\label{3.4b}
    F_0(t)=1-\rho_A(t)-\rho_B(t) \geq \rho_B(t) \, .
\end{equation}

\section{\label{dynamics}Analytical solution of the dynamics}

Here, the consequences of the dynamics defined in the previous
section will be analysed. First, the time evolution of the
densities of both species will be investigated. Equations for them
are readily obtained from the master equation (\ref{2.1}),
\begin{subequations}
\label{3.5}
\begin{eqnarray}
\frac{d}{dt} \rho_A(t)=\alpha F_1(t) \, , \label{3.5a} \\
\frac{d}{dt} \rho_B(t)=\beta F_2(t) \, , \label{3.5b}
\end{eqnarray}
\end{subequations}
where $F_1(t)$ and $F_2(t)$ are the densities of clusters with at
least two and three consecutive holes, respectively, defined in
Eq. (\ref{3.2}). Also, a hierarchy of equations for all the
marginal probabilities $F_r(t)$ is derived from the master
equation,
\begin{subequations}
\label{3.6}
\begin{eqnarray}
\frac{d}{dt}F_0(t) & = & -\alpha F_1(t)-\beta F_2(t) \, ,
\label{3.6a} \\
\frac{d}{dt}F_r(t) & = & -(\alpha+2\beta)F_{r+1}(t)-\left[ \alpha
r+\beta(r-1)\right] F_r(t)\, , \;\; r\geq 1 \, . \label{3.6b}
\end{eqnarray}
\end{subequations}
Combination of equations (\ref{3.5}) and (\ref{3.6a}) implies that
$F_0(t)+\rho_A(t)+\rho_B(t)$ is in fact an integral of motion, as
required by Eq. (\ref{3.3}). In order to solve the hierarchy of
equations (\ref{3.6b}), it is convenient to introduce the
generating function
\begin{equation}
\label{3.7} G(x,t)=\sum_{r=0}^{\infty} \frac{x^r}{r!} F_{r+1}(t)
\, ,
\end{equation}
such that
\begin{equation}
\label{3.8} F_r(t)=\left( \frac{\partial^{r-1} G(x,t)}{\partial
x^{r-1}} \right)_{x=0} \, , \qquad r\geq 1 \, .
\end{equation}
The generating function satisfies a linear first-order partial
differential equation, namely
\begin{equation}
\label{3.9}
\partial_t G(x,t)+\left[ \alpha+2\beta+(\alpha+\beta)x \right]
\partial_x G(x,t)=-\alpha G(x,t) \, ,
\end{equation}
which has to be solved with the initial condition
\begin{equation}
\label{3.10} G(x,0)=G_0(x)\equiv \sum_{r=0}^{\infty}
\frac{x^r}{r!} F_{r+1}(0) \, .
\end{equation}
By using standard techniques, it is obtained that
\begin{equation}
\label{3.11} G(x,t)=G_0\left[
-\frac{\alpha+2\beta}{\alpha+\beta}+\left(
\frac{\alpha+2\beta}{\alpha+\beta}+x \right)  e^{-(\alpha+\beta)t}
\right] e^{-\alpha t} \, .
\end{equation}
Then, from Eq. (\ref{3.8}) we derive the whole set of moments
$F_r(t)$ for $r\geq 1$,
\begin{equation}
\label{3.12} F_r(t)=G_0^{(r-1)}\left[
-\frac{\alpha+2\beta}{\alpha+\beta}+\left(
\frac{\alpha+2\beta}{\alpha+\beta} \right)  e^{-(\alpha+\beta)t}
\right] e^{-\alpha r t} e^{-\beta (r-1) t} \, ,
\end{equation}
where we have introduced the notation
\begin{equation}
\label{3.13} G_0^{(r)}(x) \equiv \frac{d^r G_0(x)}{dx^r} \, .
\end{equation}
In the long time limit $t\rightarrow\infty$, all the moments
$F_r(t)$ such that $r\geq 1$ vanish, provided that $\alpha\neq 0$.
This is easily understood: if adsorption of particles A is
possible, the system evolves until it gets stuck in a metastable
configuration such that all the holes are isolated, i.e., there
are no pairs of consecutive holes. On the other hand, if
$\alpha=0$, so that only adsorption of particles B is allowed,
$F_1(\infty)\neq 0$ while $F_r(\infty)=0$ for $r\geq 2$. In the
long time limit those configurations with at most two consecutive
empty sites are jammed since, in order to adsorb on a given site,
particles B need both of the nearest neighbours being empty.
Therefore, the limit $\alpha\rightarrow 0$ must be handled with
care in the present model, as the jammed configurations are
different for $\alpha=0$ and for $\alpha\rightarrow 0$ but not
identically null.

The evolution equations for the densities of particles $\rho_A(t)$
and $\rho_B(t)$ are obtained by substituting Eq. (\ref{3.12}) into
Eqs. (\ref{3.5}),
\begin{subequations}
\label{3.14}
\begin{eqnarray}
\frac{d}{dt} \rho_A(t) & = & \alpha \, G_0 \left[
-\frac{\alpha+2\beta}{\alpha+\beta}+\frac{\alpha+2\beta}{\alpha+\beta}
e^{-(\alpha+\beta)t} \right] e^{-\alpha t} \, , \label{3.14a} \\
\frac{d}{dt}\rho_B(t) & = & \beta \, G_0^\prime \left[
-\frac{\alpha+2\beta}{\alpha+\beta}+\frac{\alpha+2\beta}{\alpha+\beta}
e^{-(\alpha+\beta)t} \right] e^{-(2\alpha+\beta) t} \, .
\label{3.14b}
\end{eqnarray}
\end{subequations}
These equations must be integrated to get explicit expressions for
the densities $\rho_A(t)$ and $\rho_B(t)$. In order to do this, we
have to specify the initial condition for the generating function
$G_0(x)$ or, equivalently, the complete set of marginal
probabilities $F_r(0)$.

For the sake of concreteness, we will consider that the lattice is
empty at $t=0$. This is the usual initial state in adsorption
studies. Therefore,
\begin{equation}
\label{3.21} F_r(0)=1 \, , \qquad \forall r\geq 0 \, ,
\end{equation}
and, as a consequence,
\begin{equation}
\label{3.22} G_0(x)=\sum_{r=0}^{\infty} \frac{x^r}{r!}=e^x \, .
\end{equation}
From Eq.\ (\ref{3.12}) we get
\begin{equation}
\label{3.23} F_r(t)=\exp\left[-\frac{\alpha+2\beta}{\alpha+\beta}+
\frac{\alpha+2\beta}{\alpha+\beta}e^{-(\alpha+\beta)t}-\alpha r
t-\beta(r-1) t \right] \, ,
\end{equation}
for $r\geq 1$. In the long time limit $t\rightarrow\infty$,
\begin{equation}
\label{3.24} F_r(\infty)=\lim_{t\rightarrow\infty} F_r(t)=0 \, ,
\qquad \forall r\geq 1 \, ,
\end{equation}
provided that $\alpha\neq 0$, consistently with the above general
discussion. As already indicated, for $\alpha=0$ the adsorption of
particles A is impossible, and, in general, $F_1(\infty)\neq 0$.
Use of Eq. (\ref{3.22}) into Eq. (\ref{3.14}) yields
\begin{subequations}
\label{3.25}
\begin{eqnarray}
\frac{d}{dt}\rho_A(t) & = & \alpha
\exp\left[-\frac{\alpha+2\beta}{\alpha+\beta}+
\frac{\alpha+2\beta}{\alpha+\beta}e^{-(\alpha+\beta)t}-\alpha t
\right] \, , \label{3.25a} \\
\frac{d}{dt}\rho_B(t) & = & \beta
\exp\left[-\frac{\alpha+2\beta}{\alpha+\beta}+
\frac{\alpha+2\beta}{\alpha+\beta}e^{-(\alpha+\beta)t}-(\alpha+2
\beta) t \right] \, . \label{3.25b}
\end{eqnarray}
\end{subequations}
These equations can be integrated, with the result
\begin{subequations}
\label{3.27}
\begin{eqnarray}
\rho_A(t) & = & \frac{1}{1+r} e^{-\frac{1+2r}{1+r}}
\int_{e^{-(\alpha+\beta)t}}^1 du \, u^{-\frac{r}{1+r}}
e^{\frac{1+2r}{1+r} u} \, , \label{3.27a} \\
\rho_B(t) & = & \frac{r}{1+r} e^{-\frac{1+2r}{1+r}}
\int_{e^{-(\alpha+\beta)t}}^1 du \, u^{\frac{1}{1+r}}
e^{\frac{1+2r}{1+r} u} \, , \label{3.27b}
\end{eqnarray}
\end{subequations}
where we have introduced the ratio of the attempt frequencies of
adsorption
\begin{equation}
\label{3.26} r=\frac{\beta}{\alpha} \, .
\end{equation}
Integration by parts of Eq. (\ref{3.27b}) gives an explicit
relationship between both densities $\rho_A(t)$ and $\rho_B(t)$,
\begin{equation}
\label{3.28} \rho_B(t)=\frac{r}{1+2r} \left\{ 1-e^{-\alpha t}
e^{\frac{1+2r}{1+r}\left[e^{-(\alpha+\beta)t}-1\right]} -\rho_A(t)
\right\} \, .
\end{equation}

Let us analyse the long time limit of the densities, when the
system gets jammed in a metastable configuration and no more
adsorption events are possible. In this limit, the densities tend
to the values
\begin{subequations}
\label{3.29}
\begin{eqnarray}
\rho_A(\infty) & = & \frac{1}{1+r} e^{-\frac{1+2r}{1+r}}\int_0^1
du \, u^{-\frac{r}{1+r}} e^{\frac{1+2r}{1+r}u} \, , \label{3.29a}
\\
\rho_B(\infty) & = & \frac{r}{1+r} e^{-\frac{1+2r}{1+r}} \int_0^1
du \, u^{\frac{1}{1+r}} e^{\frac{1+2r}{1+r}u} \, , \label{3.29b}
\end{eqnarray}
\end{subequations}
which are related by
\begin{equation}
\label{3.30} \rho_B(\infty)=\frac{r}{1+2r} \left[ 1-\rho_A(\infty)
\right] \, ,
\end{equation}
as long as  $\alpha\neq 0$. In the long time limit, the tendency
to these asymptotic limits  are
\begin{subequations}
\label{3.30a}
\begin{eqnarray}
\rho_A(\infty)-\rho_A(t) & \sim &  e^{-\frac{1+2r}{1+r}}
e^{-\alpha t} \, , \label{3.30aa} \\
\rho_B(\infty)-\rho_B(t) & \sim & \frac{r}{2+r}
e^{-\frac{1+2r}{1+r}} e^{-(2\alpha+\beta)t} \, . \label{3.30ab}
\end{eqnarray}
\end{subequations}
Both behaviours are exponential, but the characteristic time
$(2\alpha+\beta)^{-1}$for particles B is smaller than the
characteristic time $\alpha^{-1}$ for particles A. Thus, the total
coverage of the line $\theta(t)$, defined by Eq. (\ref{3.1c}),
verifies\begin{equation}\label{3.30b}
    \theta_J-\theta(t) =
    \rho_A(\infty)-\rho_A(t)+\rho_B(\infty)-\rho_B(t) =
    e^{-\frac{1+2r}{1+r}} e^{-\alpha t}+{\cal O}\left[ e^{-(2\alpha+\beta)t}
    \right] \, ,
\end{equation}
where $\theta_J$ is the total coverage at jamming, given by Eq.
(\ref{3.1d}). The asymptotics of the coverage is dominated by the
contribution of particles A. This is easily understood, since as
time increases a configuration will be reached where no more
adsorption of particles B are possible. On the other hand,
particles A can still be adsorbed on the lattice. Our results show
that this situation will be found for times or the order of
$(2\alpha+\beta)^{-1}$, for which $\rho_B(t)$ has already decayed
to $\rho_B(\infty)$ while $\rho_A(t)$ is still evolving. This
exponential approach near the jamming limit is characteristic of
any adsorption process in a one-dimensional lattice. Power law
convergence, like Feder's $t^{-1}$ law \cite{Fe80}, arises when
continuous deposition is considered \cite{Po80,Sw81}.

In the limit $r\rightarrow\infty$ ($\alpha\ll\beta$), Eq.
(\ref{3.29b}) leads to
\begin{equation}
\label{3.31} \lim_{r\rightarrow\infty} \rho_B(\infty)=\frac{1}{2}
\left( 1-e^{-2} \right) \, ,
\end{equation}
and, therefore,
\begin{equation}
\label{3.32} \lim_{r\rightarrow\infty}
\rho_A(\infty)=1-2\lim_{r\rightarrow\infty}\rho_B(\infty)=e^{-2}
\, ,
\end{equation}
which is nonzero.  On the other hand, for $\alpha=0$ we know that
$\rho_A(t)=0$ for all $t$. This shows again the singularity of the
case $\alpha=0$. For any $\alpha\neq 0$, the system gets stuck in
configurations characterised by having all the holes isolated,
while for $\alpha=0$ those configurations having, at most, two
consecutive holes are also metastable. The adsorption of particles
A on a previously jammed configuration of particles B is analysed
in the Appendix. In that situation, the asymptotic density of
particles A, Eq. (\ref{a11}), equals Eq. (\ref{3.32}). For
$\alpha\rightarrow 0$ but nonzero, the system first reaches, over
a time scale of the order of $\beta^{-1}$, a jammed configuration
with only particles B. Afterwards, for times of the order of
$\alpha^{-1}\gg\beta^{-1}$, particles A are adsorbed on this
state, leading to an asymptotic density of particles A given by
Eq. (\ref{3.32}). The asymptotic total coverage, defined in Eq.
(\ref{3.1d}), in the limit case we are considering is
\begin{equation}
\label{3.33} \lim_{r\rightarrow\infty}
\theta_J=\lim_{r\rightarrow\infty} \left[
\rho_A(\infty)+\rho_B(\infty) \right]
=\frac{1}{2}\left(1+e^{-2}\right) \, ,
\end{equation}
which also equals Eq.\ (\ref{a12}), as expected on the basis of
the discussion above.

The limit $r\rightarrow 0$, i.e., $\beta\rightarrow 0$ does not
present any kind of singularity. For $r=0$ is is readily obtained
that
\begin{equation}
\label{3.34} \lim_{r\rightarrow 0}\rho_A(\infty)=1-e^{-1} \, ,
\qquad \lim_{r\rightarrow 0}\rho_B(\infty)=0 \, .
\end{equation}
The system only contains particles A, and the result agrees with
the one obtained for the one-component system \cite{Ev93}.

\begin{figure}
\includegraphics[scale=1]{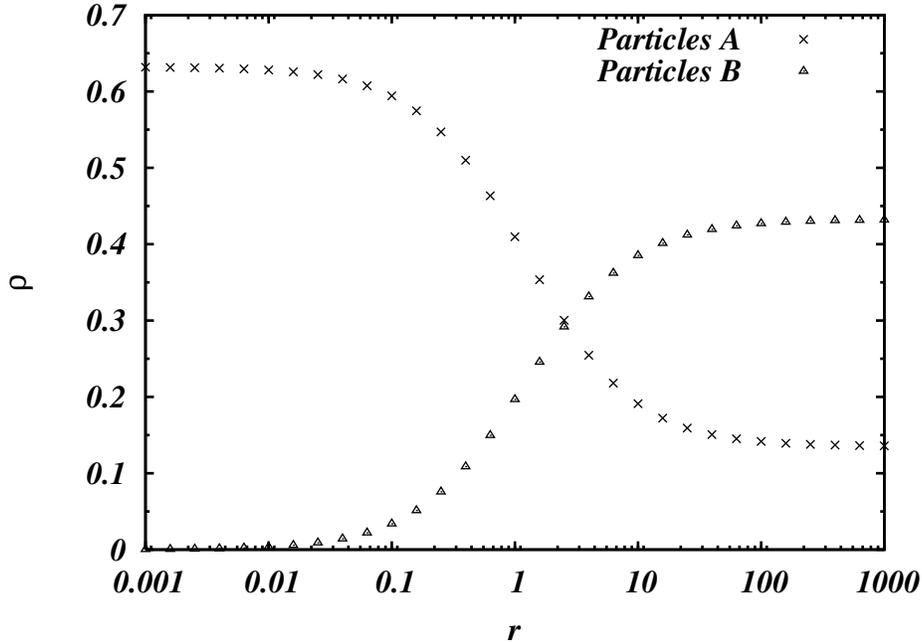}
  \caption{
Plot of the asymptotic densities $\rho_A(\infty)$ (crosses) and
$\rho_B(\infty)$ (triangles) as a function of the adsorption rates
ratio $r$. Note that $\lim_{r\rightarrow\infty}\rho_A(\infty)\neq
0$, as expressed by Eq. (\ref{3.32}).}\label{fig1}
\end{figure}

\begin{figure}
  \includegraphics[scale=1]{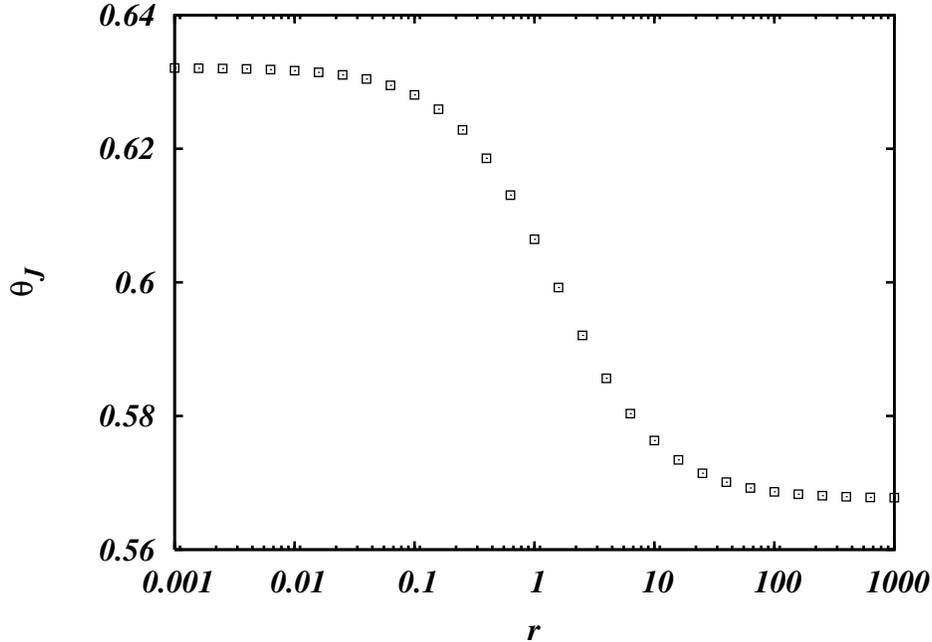}
  \caption{
Plot of the asymptotic total coverage $\theta_J$ as a function of
$r$.}\label{fig2}
\end{figure}

As it would be expected on physical grounds, the asymptotic
density of particles $\rho_B(\infty)$ is a monotonic increasing
function of $r$, while $\rho_A(\infty)$ decreases with $r$. A plot
of both asymptotic densities as a function of the adsorption rates
ratio $r$ is shown in Fig.\ \ref{fig1}. The jamming coverage can
be written in terms of $\rho_A(\infty)$ by using Eq. (\ref{3.30}),
\begin{equation}
\label{3.35}
\theta_J=\rho_A(\infty)+\rho_B(\infty)=\frac{r}{1+2r}+\frac{1+r}{1+2r}
\rho_A(\infty) \, ,
\end{equation}
which is a decreasing function of $r$, varying from
\begin{equation}
\label{3.36} \lim_{r\rightarrow 0} \theta_J=1-e^{-1} \simeq 0.63
\end{equation}
to
\begin{equation}
\label{3.37} \lim_{r\rightarrow\infty} \theta_J=\frac{1}{2} \left(
1+e^{-2}\right) \simeq 0.57 \, .
\end{equation}
Figure \ref{fig2} shows the total coverage $\theta_J$ as a
function of $r$. The maximum value of the coverage is obtained for
$r\rightarrow 0$ ($\beta\rightarrow 0$), i.e., when the adsorption
of particles B is forbidden, and there are only particles A in the
system. This is reasonable on physical grounds: due to the
facilitation rule, the domain between two particles B cannot be
completely full of particles A. Therefore, the density of
particles B is a lower bound for the density of holes. As the
density of particles B  increases with $r=\beta/\alpha$, the
asymptotic density of holes $F_0(\infty)$ also increases with $r$
and the jamming coverage $\theta_J$ decreases, since
$\theta_J=1-F_0(\infty)$. This behaviour is in contrast to all the
previous results for RSA in continuum parking models
\cite{Bo01,KSByK02a,KSByK02b} and in lattice models
\cite{Bo92,BayPr91}. In these works, it is reported that the
substrate is covered either less or more efficiently by a binary
mixture than by a single species. The model presented here,
however, leads to binary coverage lying between that of the two
monodisperse cases. This is a consequence of the cooperativity of
the adsorption processes taking place in our system.

\section{\label{corr} Correlations}

In order to get a more complete description of the steady state,
we have also investigated some correlation functions. In the
steady state, holes are isolated, i.e., surrounded by two
particles. In particular, we will be interested in the structure
of the lattice surrounding these isolated holes. In the previous
Section, the time evolution of the total hole density $F_0(t)$ has
been found, and here the density of holes for a given
configuration of their two nearest neighbours will be studied. In
the asymptotic stationary state, they can be both particles A
(AA), one particle A and one particle B (AB or BA), and two
particles B (BB). Let us define the density
\begin{equation}\label{4.1}
    \Phi_0^{AA}(t)=\langle (1-n_{i-1})m_{i-1}n_i m_i
    (1-n_{i+1})m_{i+1} \rangle_t \, ,
\end{equation}
that corresponds to the density of empty sites surrounded by two
particles A. In a similar way, we introduce
\begin{equation}\label{4.2}
    \Phi_0^{AB}(t)=\langle (1-n_{i-1})m_{i-1}n_i m_i
    n_{i+1}(1-m_{i+1}) \rangle_t \, ,
\end{equation}
\begin{equation}\label{4.3}
    \Phi_0^{BA}(t)=\langle n_{i-1}(1-m_{i-1})n_i m_i
    (1-n_{i+1})m_{i+1} \rangle_t \, ,
\end{equation}
\begin{equation}\label{4.4}
    \Phi_0^{BB}(t)=\langle n_{i-1}(1-m_{i-1})n_i m_i
    n_{i+1}(1-m_{i+1}) \rangle_t \, ,
\end{equation}
corresponding to the density of holes between a particle A on the
left and a particle B on the right ($\Phi_0^{AB}$), etc. In the
jammed state,
\begin{equation}\label{4.4b}
    F_0(\infty)=\Phi_0^{AA}(\infty)+\Phi_0^{AB}(\infty)+
    \Phi_0^{BA}(\infty)+\Phi_0^{BB}(\infty) \, .
\end{equation}
Nevertheless, an analogous expression does not hold for arbitrary
time $t$, since then it is possible to find configurations with
two adjacent holes. This is clear from the fact that the moments
$F_r(t)$, with $r\geq 1$, do not vanish except in the limit
$t\rightarrow\infty$. Then, the asymptotic fraction of holes
between two particles A at jamming is
\begin{equation}\label{4.5a}
    x_J^{AA}=
    \frac{\Phi_0^{AA}(\infty)}{F_0(\infty)} \, .
\end{equation}
Similarly, we define
\begin{equation}\label{4.5b}
    x_J^{AB}=
    \frac{\Phi_0^{AB}(\infty)}{F_0(\infty)} \, ,
\end{equation}
as the relative density of holes between a particle A to the left
and a particle B to its right,
\begin{equation}\label{4.5c}
    x_J^{BA}=
    \frac{\Phi_0^{BA}(\infty)}{F_0(\infty)} \, ,
\end{equation}
which gives the relative density of holes between a particle B to
the left and a particle A to its right, and
\begin{equation}\label{4.5d}
    x_J^{BB}=
    \frac{\Phi_0^{BB}(\infty)}{F_0(\infty)} \, ,
\end{equation}
for the fraction of holes between two particles B.

\begin{figure}
  \includegraphics[scale=1]{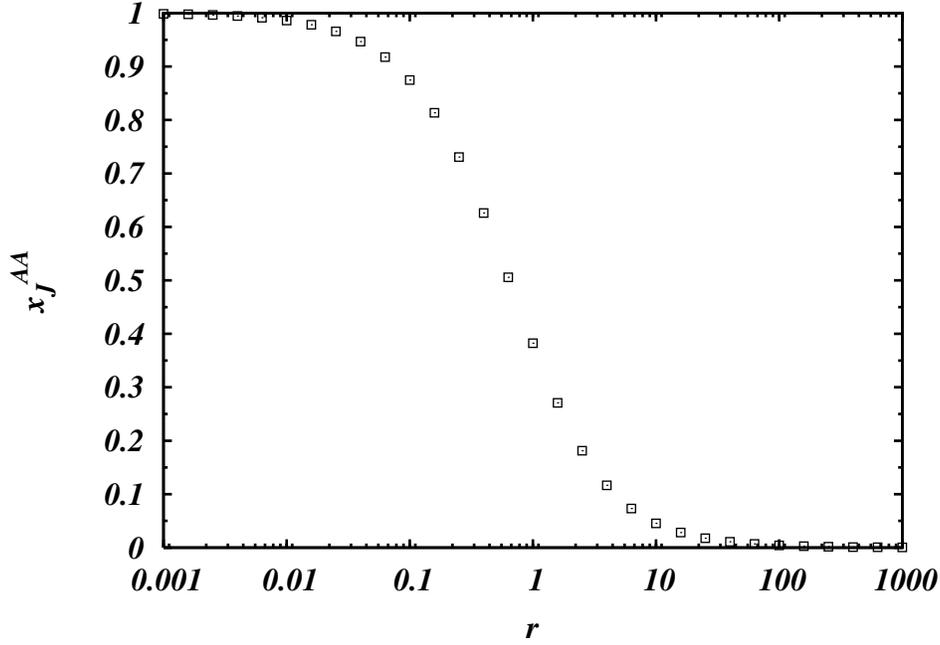}
  \caption{
Plot of the relative density $x_J^{AA}(\infty)$ (squares) as a
function of the adsorption rates ratio $r$ (in a logarithmic
scale). }\label{fig3}
\end{figure}

The evolution equations for these moments are again readily
obtained from the master equation. For the density of holes
between two particles A, one gets
\begin{equation}\label{4.6}
    \frac{d}{dt}\Phi_0^{AA}(t)=\alpha \left[
    \Phi_2^A(t)+\Phi_3^A(t) \right] \, ,
\end{equation}
where
\begin{equation}\label{4.7}
    \Phi_r^A(t)=\langle n_i m_i n_{i+1} m_{i+1} \ldots
    n_{i+r-1} m_{i+r-1} (1-n_{i+r}) m_{i+r} \rangle_t \, ,
\end{equation}
i.e., it is the probability of finding $r$ consecutive holes from
site $i$ onwards, and a particle A on site $i+r$. These
probabilities obey the hierarchy of equations
\begin{equation}\label{4.8}
    \frac{d}{dt}\Phi_r^A=-\left( \frac{\alpha}{2}+\beta \right)
    \Phi_{r+1}^A-\left[ \alpha (r-1)+\beta (r-2) \right]
    \Phi_r^A+\frac{\alpha}{2} \left( F_r+F_{r+1} \right) \, ,
\end{equation}
for $r\geq 2$. For an arbitrary initial condition, this hierarchy
of equations can be solved by a generating function method
analogous to the one used to integrate the evolution equations for
$F_r(t)$, in spite of the inhomogeneous term proportional to
$F_r+F_{r+1}$. Nonetheless, for the initially empty lattice
configuration, it is simpler to assume that
\begin{equation}\label{4.9}
    \Phi_r^A(t)=\xi_A(t) e^{-[\alpha(r-1)+\beta(r-2)]t} \, .
\end{equation}
This time dependence is suggested by the own hierarchy of
equations (\ref{4.8}) and, also, by the time dependence of the
inhomogeneous term, which is precisely of this form, as follows
from Eq. (\ref{3.23}). By substituting Eq.\ (\ref{4.9}) into Eq.\
(\ref{4.8}), a differential equation for $\xi_A(t)$ follows,
\begin{equation}\label{4.10}
    \frac{d\xi_A}{dt}=-\left( \frac{\alpha}{2}+\beta \right)
    e^{-(\alpha+\beta)t} \xi_A+\frac{\alpha}{2}
    \exp\left[ -\frac{\alpha+2\beta}{\alpha+\beta}+\frac{\alpha+2\beta}{\alpha+\beta}
    e^{-(\alpha+\beta)t}\right] e^{-(\alpha+\beta)t} \left[ 1+e^{-(\alpha+\beta)t}
    \right] \, .
\end{equation}
This is easily integrated, with the result
\begin{eqnarray}
    \xi_A(t) & = & \frac{2\alpha\beta}{(\alpha+2\beta)^2}
    \exp\left[ -\frac{\alpha+2\beta}{2(\alpha+\beta)}+\frac{\alpha+2\beta}{2(\alpha+\beta)}
    e^{-(\alpha+\beta)t} \right] \nonumber \\ & & +
    \frac{\alpha}{\alpha+2\beta}
    \exp\left[ -\frac{\alpha+2\beta}{\alpha+\beta}+\frac{\alpha+2\beta}{\alpha+\beta}
    e^{-(\alpha+\beta)t}\right]
    \left[ \frac{\alpha}{\alpha+2\beta}-e^{-(\alpha+\beta)t}
    \right] \, . \label{4.11}
\end{eqnarray}
Thus, we have identified all the moments $\Phi_r^A(t)$ and, by
using Eq.\ (\ref{4.6}), obtain the following closed evolution
equation for $\Phi_0^{AA}(t)$,
\begin{equation}\label{4.12}
    \frac{d}{dt}\phi_0^{AA}(t)=\alpha \, \xi_A(t) e^{-\alpha t}
    \left[ 1+e^{-(\alpha+\beta)t} \right] \, .
\end{equation}
Since the lattice was initially empty, $\Phi_0^{AA}(t=0)=0$, and
\begin{eqnarray}
  \Phi_0^{AA}(t) &=& \frac{2r}{(1+r)(1+2r)^2}
  e^{-\frac{1+2r}{2(1+r)}}
  \int_{e^{-(\alpha+\beta)t}}^1 du \, u^{-\frac{r}{1+r}} (1+u) e^{\frac{1+2r}{2(1+r)}u}
   \nonumber \\
  & & + \frac{1}{(1+r)(1+2r)} e^{-\frac{1+2r}{1+r}}
  \int_{e^{-(\alpha+\beta)t}}^1 du \, u^{-\frac{r}{1+r}} (1+u)
  \left( \frac{1}{1+2r}-u\right)
  e^{\frac{1+2r}{1+r}u} \, , \label{4.13}
\end{eqnarray}
where $r$ was  defined in Eq.\ (\ref{3.26}). Therefore, the
asymptotic density of holes surrounded by two particles A is
\begin{eqnarray}
  \Phi_0^{AA}(\infty) &=& \frac{2r}{(1+r)(1+2r)^2}
  e^{-\frac{1+2r}{2(1+r)}}
  \int_0^1 du \, u^{-\frac{r}{1+r}} (1+u) e^{\frac{1+2r}{2(1+r)}u}
   \nonumber \\
  & & + \frac{1}{(1+r)(1+2r)} e^{-\frac{1+2r}{1+r}}
  \int_0^1 du \, u^{-\frac{r}{1+r}} (1+u)
  \left( \frac{1}{1+2r}-u\right)
  e^{\frac{1+2r}{1+r}u} \, . \label{4.14}
\end{eqnarray}
In the limit $r\rightarrow 0$, this equation reduces to
\begin{equation}\label{4.15}
    \lim_{r\rightarrow 0} \Phi_0^{AA}(\infty)=e^{-1} \int_0^1 du
    \, (1+u)(1-u)e^u = e^{-1}=\lim_{r\rightarrow 0} F_0(\infty) \,
    .
\end{equation}
This is readily understood, as the limit $r\rightarrow 0$
corresponds to $\beta=0$, in which no adsorption events for
particles B are possible. On the other hand, in the limit
$r\rightarrow\infty$,
\begin{equation}\label{4.16}
    \lim_{r\rightarrow\infty}\Phi_0^{AA}(\infty)=0 \, .
\end{equation}
This result may appear as nontrivial, since it must be stressed
that when $r\rightarrow\infty$ there is a finite, nonvanishing,
density of particles A in the steady state, given by Eq.
(\ref{3.32}). Nevertheless, as discussed below that equation, the
adsorption of particles A takes place for very long times $t={\cal
O}(\alpha^{-1})\gg\beta^{-1}$, on sites with at most one empty
nearest neighbour. As the latter must be next to a particle B, it
follows that $\Phi_0^{AA}$ vanishes for $r\rightarrow\infty$. In
Fig.\ \ref{fig3}, the fraction of holes between two particles A
$x_J^{AA}$, defined in Eq. (\ref{4.5a}), is plotted as a function
of $r$. It decreases monotonically from unity for $r=0$ to zero
for $r\rightarrow\infty$, as expected.

\begin{figure}
  \includegraphics[scale=1]{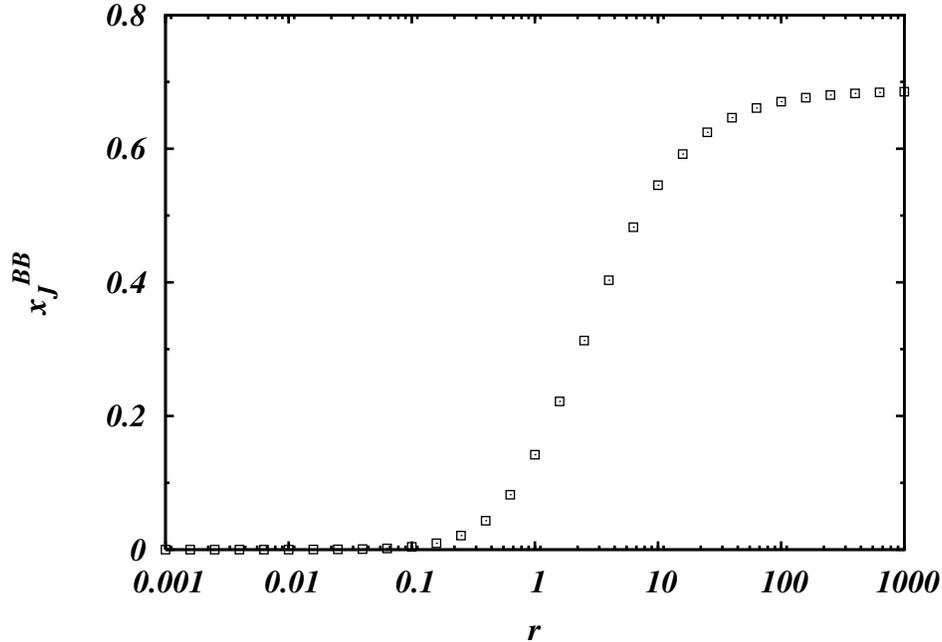}
  \caption{
Plot of the concentration $x_J^{BB}$ as a function of $r$ (in a
logarithmic scale). Note that $x_J^{BB}$ does not tend to unity
for $r\rightarrow\infty$.}\label{fig4}
\end{figure}

Let us now analyse the density of holes between two particles B,
$\Phi_0^{BB}(t)$. From the master equation, it is obtained
\begin{equation}\label{4.17}
    \frac{d}{dt} \Phi_0^{BB}=2\beta\Phi_3^B \, ,
\end{equation}
where now
\begin{equation}\label{4.18}
    \Phi_r^B(t)=\langle n_i m_i n_{i+1} m_{i+1} \ldots
    n_{i+r-1} m_{i+r-1} n_{i+r} (1-m_{i+r}) \rangle_t \, ,
\end{equation}
i.e., it is the probability of finding $r$ consecutive holes from
site $i$ onwards, and a particle B on site $i+r$. Note the analogy
of the notation used here with the one introduced for the moments
$\Phi_r^A$ defined in Eq.\ (\ref{4.7}). These moments obey the
hierarchy
\begin{equation}\label{4.19}
    \frac{d}{dt}\Phi_r^B=-\left( \frac{\alpha}{2}+\beta \right)
    \Phi_{r+1}^B-\left[ \alpha(r-1)+\beta(r-2)\right]
    \Phi_r^B+\beta F_{r+1} \, ,
\end{equation}
for $r\geq 2$. Again, for the initially empty lattice we are
considering, the family of solutions
\begin{equation}\label{4.20}
    \Phi_r^B(t)=\xi_B(t) e^{-[\alpha(r-1)+\beta(r-2)]t} \,
\end{equation}
can be considered. Substitution of the above expression into Eq.\
(\ref{4.19}) yields a closed differential equation for $\xi_B(t)$,
namely
\begin{equation}\label{4.21}
    \frac{d}{dt}\xi_B=-\left(\frac{\alpha}{2}+\beta\right)e^{-(\alpha+\beta)t}
    \xi_B+\beta \exp\left[ -\frac{\alpha+2\beta}{\alpha+\beta}+
    \frac{\alpha+2\beta}{\alpha+\beta} e^{-(\alpha+\beta)t}\right]
    e^{-2(\alpha+\beta)t} \, ,
\end{equation}
which can be integrated by standard techniques, obtaining
\begin{eqnarray}\label{4.22}
    \xi_B(t) & = & \frac{4\beta(\alpha+\beta)}{(\alpha+2\beta)^2}
    \exp\left[ -\frac{\alpha+2\beta}{\alpha+\beta}+
    \frac{\alpha+2\beta}{\alpha+\beta} e^{-(\alpha+\beta)t}\right]
    \left[ 1-\frac{\alpha+2\beta}{2(\alpha+\beta)}
    e^{-(\alpha+\beta)t} \right]  \nonumber \\
    & & -\frac{2\alpha\beta}{(\alpha+2\beta)^2}
    \exp\left[ -\frac{\alpha+2\beta}{2(\alpha+\beta)}+
    \frac{\alpha+2\beta}{2(\alpha+\beta)}
    e^{-(\alpha+\beta)t}\right]\, .
    \label{4.23}
\end{eqnarray}
This provides the expressions for all the densities $\Phi_r^B$
($r\geq 2$), from Eq.\ (\ref{4.20}). Taking into account Eq.\
(\ref{4.17}) and the initial condition, it is found
\begin{eqnarray}
  \Phi_0^{BB}(t) &=& \frac{8r^2}{(1+2r)^2} e^{-\frac{1+2r}{1+r}}
  \int_{e^{-(\alpha+\beta)t}}^1 du \, u^{\frac{1}{1+r}}
  \left[ 1- \frac{1+2r}{2(1+r)} u\right] e^{\frac{1+2r}{1+r}u}
\nonumber   \\
   & & -\frac{4r^2}{(1+r)(1+2r)^2} e^{-\frac{1+2r}{2(1+r)}}
   \int_{e^{-(\alpha+\beta)t}}^1 du \, u^{\frac{1}{1+r}}
   e^{\frac{1+2r}{2(1+r)}u} \label{4.24} \, .
\end{eqnarray}
In the long time limit,
\begin{eqnarray}
  \Phi_0^{BB}(\infty) &=& \frac{8r^2}{(1+2r)^2} e^{-\frac{1+2r}{1+r}}
  \int_0^1 du \, u^{\frac{1}{1+r}}
  \left[ 1- \frac{1+2r}{2(1+r)} u\right] e^{\frac{1+2r}{1+r}u}
\nonumber   \\
   & & -\frac{4r^2}{(1+r)(1+2r)^2} e^{-\frac{1+2r}{2(1+r)}}
   \int_0^1 du \, u^{\frac{1}{1+r}}
   e^{\frac{1+2r}{2(1+r)}u} \label{4.25} \, .
\end{eqnarray}
If the adsorption of particles B is forbidden, $\beta\rightarrow
0$ or $r\rightarrow 0$, the expected result,
\begin{equation}\label{4.26}
    \lim_{r\rightarrow 0}\Phi_0^{BB}(\infty)=0 \, ,
\end{equation}
is obtained. On the other hand, when no adsorption of particles A,
$\alpha\rightarrow 0$ or $r\rightarrow\infty$,
\begin{equation}\label{4.27}
    \lim_{r\rightarrow\infty} \Phi_0^{BB}(\infty)=2e^{-2}\int_0^1
    du (1-u) e^{2u}=\frac{1}{2}-\frac{3}{2}e^{-2} \, .
\end{equation}
Then, $\Phi_0^{BB}(\infty)$ does not equal the total density of
holes $F_0(\infty)$ in this limit. This is due to the fact that
for $\alpha\neq 0$, although very small, there is a non vanishing
density of particles A, given by Eq. (\ref{3.32}). The consistent
result is
\begin{equation}\label{4.28}
    \lim_{r\rightarrow\infty} \left[
    F_0(\infty)-\Phi_0^{BB}(\infty) \right]
    =\lim_{r\rightarrow\infty} \rho_A(\infty) \, .
\end{equation}
As we have already discussed above, in the limit
$r\rightarrow\infty$ particles A are adsorbed once the system has
reached a jammed configuration in which only particles B have been
adsorbed ($\alpha=0$). These intermediate configurations are
characterised by having clusters of holes involving at most two
consecutive sites. Therefore, particles A will be adsorbed in one
of the empty sites of those clusters with two adjacent holes,
i.e.,
\begin{equation}\label{4.29}
    \lim_{r\rightarrow\infty} \left[\Phi_0^{BB}(\infty)+\rho_A(\infty) \right]
    =\lim_{r\rightarrow\infty} F_0(\infty)  \, ,
\end{equation}
because $\rho_A(\infty)$ equals the density of holes between a
particle B and another hole in the intermediate ``metastable''
state. In Fig. \ref{fig4}, the fraction of holes between two
particles B, $x_J^{BB}$, is plotted as a function of the
adsorption rates ratio $r$. It is seen that $x_J^{BB}$ is a
monotonically increasing function of $r$. It must be noted that
$\lim_{r\rightarrow\infty}x_0^{BB}\neq 1$, since in that limit
$F_0^{BB}(\infty)$ does not equal the total density of holes
$F_0(\infty)$, as expressed by Eq. (\ref{4.29}). More concretely,
\begin{equation}\label{4.29b}
    \lim_{r\rightarrow\infty}x_J^{BB}=\lim_{r\rightarrow\infty}
    \frac{F_0^{BB}(\infty)}{F_0(\infty)}=\frac{1-3e^{-2}}{1-e^{-2}}
    \, .
\end{equation}

\begin{figure}
  \includegraphics[scale=1]{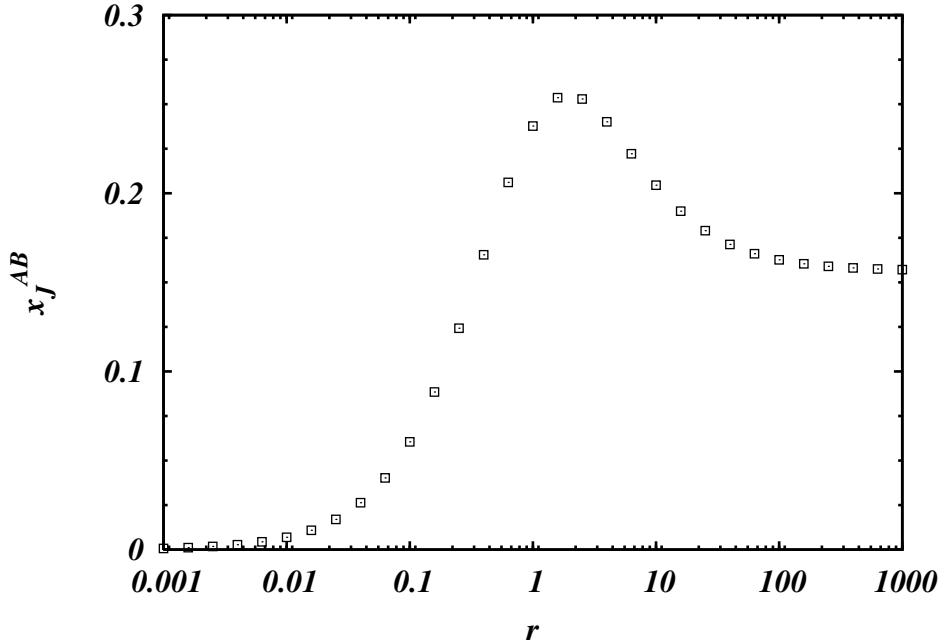}
  \caption{
Plot of $x_J^{AB}(\infty)$ as a function of $r$ (in a logarithmic
scale). A maximum occurs for $r\simeq 2$.}\label{fig5}
\end{figure}

Taking into account Eq. (\ref{4.4b}), it is possible to obtain
also the density of holes between one particle A and one particle
B, independently of their relative positions
\begin{equation}\label{4.30}
    \Phi_0^{AB}(\infty)=\Phi_0^{BA}(\infty)=\frac{1}{2}\left[F_0(\infty)
    -\Phi_0^{AA}(\infty)-\Phi_0^{BB}(\infty)\right] \, .
\end{equation}
The fraction of holes between two distinct particles
$x_J^{AB}=x_J^{BA}$ is plotted in Fig.\ \ref{fig5}. It has a
non-monotonic behaviour, exhibiting a maximum for $r\simeq 2$,
i.e., in the region where the attempt rates of adsorption $\alpha$
and $\beta$ are of the same order of magnitude. For $r\rightarrow
0$, $x_J^{AB}\rightarrow 0$, since no particles B are adsorbed on
the lattice. On the other hand, in the limit $r\rightarrow\infty$
a ``minimum'' adsorption of particles A is required on those
groups of two consecutive holes left by the previous adsorption of
particles B, as already discussed. Then,
\begin{equation}\label{4.31}
    \lim_{r\rightarrow\infty} x_J^{AB}=
    \lim_{r\rightarrow\infty} \frac{\Phi_0^{AB}(\infty)}{F_0(\infty)}=
    \lim_{r\rightarrow\infty} \frac{\rho_A(\infty)}{2F_0(\infty)}=
    \frac{e^{-2}}{1-e^{-2}} \, ,
\end{equation}
since in that limit $\Phi_0^{AB}+\Phi_0^{BA}=2\Phi_0^{AB}$ should
equal $\rho_A$, as expressed by Eqs. (\ref{4.28}) or (\ref{4.29}).

\section{\label{conc}Conclusions}

The objective here has been to study a one-dimensional model for
the adsorption of a binary mixture on a lattice with nearest
neighbour cooperative effects. Both species are of the same size
(monomers), but they differ in the strength of the NN repulsive
interactions, as measured by the number of NN empty sites required
for a particle to adsorb. Emphasis has been put on the jammed
configuration reached by the system in the long time limit. The
asymptotic  coverage of the lattice by the mixture differs from
previous results both for the random sequential adsorption of
mixtures on lattices and continuous substrates
\cite{BayPr91,Bo92,Bo01,KSByK02a,KSByK02b}. The mixture always
covers the lattice more efficiently than the species of particles
feeling stronger NN repulsion, and less efficiently than the
species experiencing a weaker one. In fact, the jamming coverage
is a monotonic function of the ratio between the attempt rates of
adsorption for the two species. This is a consequence of the
cooperativity of the adsorption events considered in our model.

In the long time limit, the tendency to the asymptotic value of
the coverage is dominated by the least cooperative species. This
is similar to the behaviour observed in models with species of
different sizes, in which this approach is dominated by the
smaller species  \cite{BayPr91,Bo92,Bo01,KSByK02a,KSByK02b}.
Nevertheless, in our model the approach is exponential and,
therefore, Feder's $t^{-1}$ power law \cite{Fe80} does not hold.
This is a consistent result, since exponential behaviour is
characteristic of any d-dimensional lattice deposition, evolving
to a power law when a continuum deposition limit is introduced
\cite{Po80,Sw81}.

The structure of the jamming configuration has also been
investigated by computing some spatial correlation functions.
Here, the interest has been put on the nature of the holes, i.e.,
the kind of particles surrounding them. The relative densities of
holes between two identical particles, of either of the two
species, show a monotonic behaviour as a function of the
adsorption rates ratio. On the other hand,  the fraction of holes
between two particles of different species displays a maximum in
the region where both attempt rates of adsorption are of the same
order of magnitude. This may be a relevant information in order to
determine the properties of the adsorbed layer in a given physical
system.

The model discussed here is closely related to the so-called
facilitated Ising models, first introduced by Fredrickson and
Andersen \cite{FyA84,FyA85,FyB86}. The main characteristic of
these systems is that, although their thermal equilibrium
properties are trivial, their dynamics is highly cooperative and
very slow. A recent review on these kinetically constrained Ising
models can be found in Ref.\ \cite{RyS03}. They are appropriate
and have been used to model physical systems as structural glasses
\cite{FyB86,ByH91a,ByH91b,RyJ95} or dense granulates
\cite{BPyS99,BPyS00,LyD01,BFyS02,SGyL02,PyB02}, in which the
structural rearrangement of a given region may be slowed down or
even blocked by the configuration of the surroundings.

Irreversible cooperative adsorption processes have also been used
as models for the free relaxation between two shakes in vibration
experiments with granular systems
\cite{BPyS99,BPyS00,LyD01,BFyS02,SGyL02,PyB02}. Let us consider an
horizontal section of a real granular binary mixture, near the
bottom of its container. During the free evolution of the system,
i.e., only under the action of gravity, particles can only go
down, as long as there is enough empty space in their
surroundings.  The total density of particles in the layer grows
until the hard-core interaction prevents more movements of
particles, and a \textit{metastable} (mechanically stable)
configuration is reached.  Different kind of grains may need also
different amounts of free space in their surroundings to be
adsorbed on or desorbed from the layer. In our system, these
kinetic constraints are modeled by the different ``facilitation
rules'' for the adsorption of both species. On the other hand,
when the system is submitted to vertical vibration, particles can
go up, decreasing the density in the layer. This vibration
dynamics can be modeled by allowing desorption events. In this
way, a simple model to analyse granular segregation phenomena in a
binary mixture can be formulated \cite{PyB03}.

\begin{acknowledgments}

This research was supported by the Ministerio de Educaci\'on y
Ciencia (Spain) through Grant No.\ FIS2005-01398 (partially
financed by FEDER funds).

\end{acknowledgments}

\appendix

\section{\label{big}Adsorption of particles A on a jammed configuration of particles B}

In this Appendix, we will briefly analyse  the dynamics of a
system in which particles A adsorb on a jammed state corresponding
to a previous adsorption of particles B. The adsorption of B
particles is equivalent to the one-dimensional RSA with NN
exclusion or blocking, whose solution is well-known \cite{Ev93}.
In the jammed state of particles B, the density of holes reads
\begin{equation}\label{a1}
    F_0(\infty)=\frac{1}{2}\left( 1+e^{-2}\right) \, ,
\end{equation}
and the density of pairs of consecutive holes is
\begin{equation}\label{a2}
    F_1(\infty)=e^{-2} \, .
\end{equation}
Since the jammed state of particles B is characterised by having
at most two consecutive holes,
\begin{equation}\label{a3}
    F_r(\infty)=0 \, \qquad \text{for all $r\geq 2$.}
\end{equation}
The asymptotic density of particles B directly follows from the
density of holes,
\begin{equation}\label{a4}
    \rho_B(\infty)=1-F_0(\infty)=\frac{1}{2}\left( 1-e^{-2}\right)
    \, .
\end{equation}

Let us now consider that this jammed state for the system of
particles B is the initial state for the adsorption of particles
A. Therefore, in terms of the moments $F_r(t)$ we will have
\begin{equation}
\label{a5} F_r(0)=0 \, , \forall r\geq 2   \, ; \qquad F_1(0)\neq
0 \, , \qquad F_0(0)\neq 0 \, ,
\end{equation}
being $F_0(0)$ and $F_1(0)$ the initial density of holes and of
pairs of consecutive holes, given by Eqs.\ (\ref{a1}) and
({\ref{a2}), respectively. Besides, the initial densities of
particles are
\begin{equation}
\label{a6} \rho_A(0)=0 \, , \qquad \rho_B(0)\neq 0 \, ,
\end{equation}
where $\rho_B(0)$ is given by Eq.\ (\ref{a4}).

In order to solve this problem, the formalism developed for the
complete system in Sec.\ \ref{dynamics}, making $\beta=0$, can be
used. The initial generating function is,
\begin{equation}
\label{a7} G_0(x)=\sum_{r=0}^\infty \frac{x^r}{r!}
F_{r+1}(0)=F_1(0) \, ,
\end{equation}
which is independent of $x$. Then, Eq. (\ref{3.12}) yields
\begin{equation}
\label{a8} F_1(t)=F_1(0) e^{-\alpha t} \, ,
\end{equation}
i.e., the density of pairs of consecutive holes decays
exponentially. Of course, $F_r(t)=0$ for all $r\geq 2$. The
evolution equations for the densities (\ref{3.14}) are very
simple,
\begin{equation}
\label{a9} \frac{d}{dt} \rho_B(t)=0 \, , \qquad \frac{d}{dt}
\rho_A(t)=\alpha F_1(0) e^{-\alpha t} \, ,
\end{equation}
leading to
\begin{equation}
\label{a10} \rho_A(t)=F_1(0) \left( 1-e^{-\alpha t} \right) \, .
\end{equation}
This result reflects that the initial configuration is a jammed
state of the particles B and, therefore, the holes are either
isolated or in groups of two consecutive holes, whose density is
$F_1(0)$. Particles A can be adsorbed on any of the $2 F_1(t)$
sites which are next to a hole, with a rate $\alpha/2$. As all of
these processes are independent, the density of hole pairs
$F_1(t)$ decays exponentially as $\exp(-\alpha t)$, and the
density of particles A also increases exponentially until it
reaches its steady value
\begin{equation}\label{a11}
\rho_A(\infty)=F_1(0)=e^{-2} \, ,
\end{equation}
with a characteristic time $\tau=\alpha^{-1}$. Note that this is
precisely the limit of $\rho_A(\infty)$ when
$r=\beta/\alpha\rightarrow\infty$, Eq.\ (\ref{3.32}). Besides, the
total coverage of the line will be
\begin{equation}\label{a12}
    \theta_J=\rho_A(\infty)+\rho_B(\infty)=\frac{1}{2}\left( 1+e^{-2}
    \right) \, ,
\end{equation}
which agrees with Eq. (\ref{3.33}).

\bibliography{references,myarticles}

\end{document}